\documentclass[sigconf]{acmart}

\usepackage{multirow}
\usepackage{tabularx}
\usepackage{}
\AtBeginDocument{%
  \providecommand\BibTeX{{%
    \normalfont B\kern-0.5em{\scshape i\kern-0.25em b}\kern-0.8em\TeX}}}

\setcopyright{acmcopyright}
\copyrightyear{2018}
\acmYear{2018}
\acmDOI{10.1145/1122445.1122456}

\acmConference[Woodstock '18]{Woodstock '18: ACM Symposium on Neural
  Gaze Detection}{June 03--05, 2018}{Woodstock, NY}
\acmBooktitle{Woodstock '18: ACM Symposium on Neural Gaze Detection,
  June 03--05, 2018, Woodstock, NY}
\acmPrice{15.00}
\acmISBN{978-1-4503-9999-9/18/06}

\begin{document}

%%
%% The "title" command has an optional parameter,
%% allowing the author to define a "short title" to be used in page headers.
% \title{Capture Long Term Dependence for Sequential Recommendation with the Help of Auxiliary Task }
\title{Learning to Structure Long-term Dependence for Sequential Recommendation}

%%
%% The "author" command and its associated commands are used to define
%% the authors and their affiliations.
%% Of note is the shared affiliation of the first two authors, and the
%% "authornote" and "authornotemark" commands
%% used to denote shared contribution to the research.

\author{Renqin Cai}
\email{rc7ne@virginia.edu}
\affiliation{%
  \institution{University of Virginia}
  \country{USA}
}

\author{Qinglei Wang}
\email{qinglei.wang@bytedance.com}
\affiliation{%
  \institution{Bytedance}
  \country{China}
}

\author{Chong Wang}
\email{chong.wang@bytedance.com}
\affiliation{%
  \institution{Bytedance}
  \country{USA}
}

\author{Xiaobing Liu}
\email{xiaobing.liu@bytedance.com}
\affiliation{%
  \institution{Bytedance}
  \country{USA}
}

%%
%% By default, the full list of authors will be used in the page
%% headers. Often, this list is too long, and will overlap
%% other information printed in the page headers. This command allows
%% the author to define a more concise list
%% of authors' names for this purpose.

%%
%% The abstract is a short summary of the work to be presented in the
%% article.
\begin{abstract}
Sequential recommendation recommends items based on sequences of users' historical actions. The key challenge in it is how to effectively model the influence from distant actions to the action to be predicted, i.e., recognizing the long-term dependence structure; and it remains an underexplored problem. 
To better model the long-term dependence structure, we propose a GatedLongRec solution in this work. 
To account for the long-term dependence, GatedLongRec extracts distant actions of top-$k$ related categories to the user's ongoing intent with a top-$k$ gating network, and utilizes a long-term encoder to encode the transition patterns among these identified actions. 
As user intent is not directly observable, we take advantage of available side-information about the actions, i.e., the category of their associated items, to infer the intents. End-to-end training is performed to estimate the intent representation and predict the next action for sequential recommendation.
Extensive experiments on two large datasets show that the proposed solution can recognize the structure of long-term dependence, thus greatly improving the sequential recommendation.

% take advantage of the category information of the target action to retrieve relevant transition patterns for improving target action prediction. Specifically, we use a gating network to predict the category of the target action. With the predicted category, we then distill the long-range historical actions to capture long-term dependence with a long-term encoder. Meanwhile, we use a short-term encoder to capture the dependence among short-range actions (short-term dependence). Both long-term and short-term dependence are used to recommend future items. Extensive experiments on two large datasets show that the proposed long-term dependence modeling solution greatly improves the sequential recommendation. 

% Due to the diversity of influence from long-range actions, 
%blindly leveraging transition patterns on long-range actions may not capture the long-term dependence accurately, 
% it is important to recognize the relevant parts of those patterns; otherwise, it might have adversarial effects on predicting the user's future actions. 

%distill the long-term dependence dynamically to improve the target action prediction. We 

%Moreover, we assume the long-term dependence of the target action is on some long-range actions over which the transition pattern is relevant. 

% which can help capture the dependence on long-range actions accurately

% to learn to distill the dependence of the target action on the long-range actions in regarding to the target action. 

% which prevents sequential recommendation from achieving better performance.
\end{abstract}

%%
%% The code below is generated by the tool at http://dl.acm.org/ccs.cfm.
%% Please copy and paste the code instead of the example below.
%%
% \begin{CCSXML}
% <ccs2012>
%  <concept>
%   <concept_id>10010520.10010553.10010562</concept_id>
%   <concept_desc>Computer systems organization~Embedded systems</concept_desc>
%   <concept_significance>500</concept_significance>
%  </concept>
%  <concept>
%   <concept_id>10010520.10010575.10010755</concept_id>
%   <concept_desc>Computer systems organization~Redundancy</concept_desc>
%   <concept_significance>300</concept_significance>
%  </concept>
%  <concept>
%   <concept_id>10010520.10010553.10010554</concept_id>
%   <concept_desc>Computer systems organization~Robotics</concept_desc>
%   <concept_significance>100</concept_significance>
%  </concept>
%  <concept>
%   <concept_id>10003033.10003083.10003095</concept_id>
%   <concept_desc>Networks~Network reliability</concept_desc>
%   <concept_significance>100</concept_significance>
%  </concept>
% </ccs2012>
% \end{CCSXML}

% \ccsdesc[500]{Computer systems organization~Embedded systems}
% \ccsdesc[300]{Computer systems organization~Redundancy}
% \ccsdesc{Computer systems organization~Robotics}
% \ccsdesc[100]{Networks~Network reliability}

%%
%% Keywords. The author(s) should pick words that accurately describe
%% the work being presented. Separate the keywords with commas.
\keywords{Sequential recommendation, neural networks, long-term dependence modeling}

%% A "teaser" image appears between the author and affiliation
%% information and the body of the document, and typically spans the
%% page.

%%
%% This command processes the author and affiliation and title
%% information and builds the first part of the formatted document.
\maketitle

\section{Introduction}

Users interact with online service platforms, such as e-commerce or news media websites to fulfill their information need \cite{palovics2014exploiting, wu2017recurrent, wang2016coevolutionary}. The series of interactions often reveal the development of users' interest~\cite{he2016fusing, wang2019recurrent}; and thus properly modeling of such data is important for accurate user understanding and service utility optimization. This has motivated the research of sequential recommendation, which makes recommendations based on the modeling of sequences of users' interactive actions with online service systems~\cite{tang2018personalized, chen2018sequential, tang2019towards, you2019hierarchical, an2019neural}.

% i.e., sequential recommendation 
%To understand the dynamics of users' interest, sequential recommendation models actions in a sequential fashion as how users generate them. 
%To satisfy users' dynamic needs, sequential recommendation models transition patterns among user actions to capture user's interest. They recommend a list of items based on a user's sequential action history~

\begin{figure*}[t]
\centering
\includegraphics[width=0.95\textwidth]{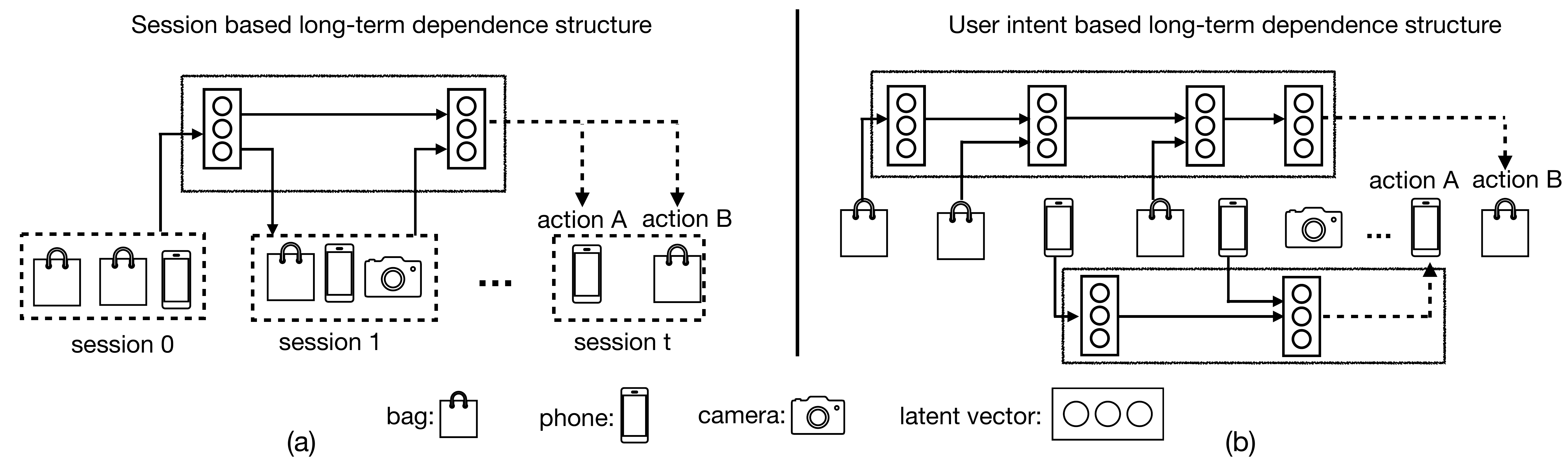} 
\caption{An example sequence of actions for sequential recommendation. The icons at the bottom represent the categories of items associated with each action. Different items of the same category is represented with the same icon. (a) visualizes the session-based long-term dependence structure, and (b) visualizes the user intent based long-term dependence structure proposed in this paper.} 
\label{fig:Sequence_example_intro}
\end{figure*}

% \begin{figure*}[t]
% \vspace{-2mm}
% \centering
% \begin{tabular}{c c}
% \includegraphics[width=0.45\textwidth]{resource/intro_example_3.pdf} & 
% \includegraphics[width=0.45\textwidth]{resource/intro_example_2.pdf} \hspace{0.2 in} \\
% \small{(a)} & \small{(b)}
% % (a) tune $\beta_m$ with fixed $\beta_s$ & & (b) tune $\beta_s$ with fixed $\beta_m$
% \end{tabular}
% \vspace{-2mm}
% \caption{A sequence of actions generated by a certain user. The icon represents the category of the item associated with an action. Different items of the same category is represented with the same icon. While (a) represents the structure of long-term dependence correlated with users' intention, (b) represents the hierarchical structure of long-term dependence agnostic to users' intention. }
%  \label{fig:Sequence_example_intro}
% \vspace{-3mm}
% \end{figure*}

Although previous work~\cite{belletti2019quantifying, jing2017neural} validates the influence from distant historical actions on the current action to be predicted, zooming into details of this type of dependence and properly structuring it are still underexplored in sequential recommendation. Following the notions defined in \cite{belletti2019quantifying}, the distant actions are typically tens or even hundreds of steps before the action to be predicted, while the recent actions are only several steps away. As Cai et al. found in \cite{cai2018modeling}, the influence from those distant actions to the current action differs from that from recent actions, i.e., long-term v.s., short-term dependence. Moreover, because the long-term dependence is accumulated over a larger number of historical actions, its structure inherently is more complex than what short-term dependence might have. This makes the modeling of such a structure non-trivial. Blindly feeding the whole sequence of actions into either a vanilla recurrent neural network (RNN) or self-attention model \cite{liu2015multi, belletti2019quantifying} can hardly recognize the subtle structure, thus leads to inaccurate modeling.

% it inherently has a more  structure than what short-term dependence might have.
% For example, under different context, different parts of long-range actions influence the current action. 

To address the limitation of existing vanilla solutions for long-term dependence modeling, several recent studies introduce hierarchical structures for fine-grained modeling of such dependence. \citet{quadrana2017personalizing} proposed a hierarchical recurrent neural network model: one RNN models transition among actions within a time-based session; and on top of it, another RNN models the transition across sessions. As a result, they explicitly structure the long-term dependency with accordance to the session boundaries. \citet{you2019hierarchical} proposed a very similar session-based hierarchical model, i.e., Hierarchical Temporal Convolutional Networks, by replacing the in-session RNN with a Temporal Convolutional Network for fast training. Improved recommendation quality was reported in both models, compared with previous vanilla RNN-based solutions. 
However, although the distant actions are structured into sessions, they are summarized into one latent vector, which is shared by all actions in the current session. This fails to differentiate a user's potentially different intents behind the actions to be performed. Take the situation illustrated in Figure \ref{fig:Sequence_example_intro} (b) as an example. All actions in the current session share the same latent state summarized from previous sessions, though the user's information need differs behind the two actions, e.g., shopping for a cellphone v.s., a handbag. This invariant long-term dependence encoding leads to indiscriminate and sub-optimal recommendations at action ``A'' and ``B'', thus undermines the utility of sequential recommendation.

As a user's sequential behaviors are driven by her/his underlying intent, the structure of long-term dependence becomes more evident when we can infer the intent driving the current action. %The intent allows to differentiate the influence of long-range actions for the long-term dependence. 
Take Figure~\ref{fig:Sequence_example_intro} (a) as an example, when the user intends to research properties of different cellphones at action ``A'', her/his past actions associated with cellphones suggest what he/she might be interested in at the moment, e.g., examine unexplored cellphones v.s., revisit previously examined ones. Unfortunately, because a user's intent is unobserved, predicting it is no easier than learning the long-term dependence structure. 

% past actions in these work is independent to the user's intention at the target action. Consequently, the long-term dependence of the action ``A" is the same as that of the action ``B", though the intention for phones at the action ``A" is different from that for bags at the action ``B".

% It treats past actions within the session as short-range actions and those from historical sessions as long-range actions. The long-term dependence is modeled with a hierarchical structure. Likewise, \citet{you2019hierarchical} proposed a hierarchical model that replaces within-session RNN with Temporal Convolutional Network (TCN) for fast training. Although the long-term dependence has been shown to help these methods improve the recommendation quality, they still do not explicitly differentiate relevant transition patterns from irrelevant ones regarding to the target action. Because of the mixed diverse transition patterns,  long-term dependence becomes harder to recognize for improving recommendation performance.
% \hspace{-0.2 in}

To break this paradox, we look for proxies of a user's intent behind the sequential actions, and construct the long-term dependence structure with respect to the proxy signal. In this paper, we choose category of the item associated with each action as the proxy, for two main reasons. First, as shown in \cite{zhu2018learning},
the category of an item chosen by a user closely correlates with the user's underlying intent. Second, category information is widely available in many recommendation scenarios, such as news and production recommendations. This ensures the generality of our solution. 

With this approximated user intent reflected on the category-level, we develop a neural sequential recommendation solution, named GatedLongRec, to structure the long-term dependence. Empirical evidence in \cite{lo2016understanding} suggests that recent actions tend to share similar intents, and therefore are well correlated with each other. Hence we employ a RNN-based intent encoder taking categories of recent actions as input to generate a representation of the user's on-going intent of the action to be predicted. Then we use this intent representation to structure the long-term dependence. Considering the uncertainty of the inferred user intent, which is introduced by using category as supervision, we propose a top-k gating network which extracts distant actions from the top-$k$ similar categories to the user's predicted intent. With actions of top-$k$ categories as input, another RNN-based long-term encoder produces $k$ per-category summary vectors for predicting the next action collectively.

In addition to the modeled long-term dependency, short-term dependence is also critical in sequential recommendation \cite{sarwar2001item, li2017neural, hidasi2015session}. We encode recent actions via a separate short-term encoder to capture this dependence in GatedLongRec. 
For each of the $k$ long-term representations, we concatenate it with the short-term dependence representation to rank all candidate items. To reflect the confidence of the learnt user's intent, We merge these $k$ recommendation lists by their corresponding attention weight obtained in the top-k gating network. 

To verify the effectiveness of our long-term dependence modeling for sequential recommendation, we conduct extensive experiments on two large datasets. Compared with state-of-art baselines, our model captures the long-term dependence more accurately, thus providing more accurate recommendations of the target actions. Our ablation analysis demonstrates the contribution of components in GatedLongRec to the improved recommendations. Besides, we analyze how the hyper-parameters influence GatedLongRec to gain some insight on the properties of GatedLongRec.

\section{Related Work}
% Sequential Recommendation is an important branch of recommendation systems, which recommends items to users based on sequences of users' actions. It leverages the sequential transition patterns among users' action history to infer users' interest, and then make recommendations in regard to the interest. Sequential transition patterns have been shown to enhance the recommendation quality. A number of research have been devoted to sequential recommendation. 

Sequential recommendation is an important direction in recommendation research, which recommends items to users based on sequences of users' historical actions. Different types of solutions have been developed for sequential recommendation, and we organize our related work discussion accordingly. %In addition, as GatedLongRec utilizes hard attention when structuring the long-term dependence and predicting the item for the target action, we include some discussion of the hard attention.

% \textbf{add some discussion about why sequential recommendation. why Long term dependence is difficult.}

Long-term dependence has been widely conjectured to present in users' sequential behaviors. From the perspective of cognitive psychology, human's memory mechanism explains how users' future behaviors are affected by distant behaviors~\cite{ryan2008hippocampal}. From the perspective of computer science, ~\citet{belletti2019quantifying} proposed a principled estimation procedure to computationally quantify the long-range dependence. These work support that better accounting for long-term dependence enables better behavioral predictions.

Fixed- or varying-order Markov models are one of the earlier attempts~\cite{rendle2010factorizing, begleiter2004prediction}, which assume the prediction depends on the last several actions. Particularly ~\citet{rendle2010factorizing} proposed a Factorizing Personalized
Markov Chains (FPMC) model for sequential recommendation which combines both the machine factorization and Markov models. Because of the introduced Markov dependence, FPMC achieves better performance than vallina matrix factorization based models. On the other hand, the Markov dependence also limits the performance of these models. Since the state-space grows exponentially with respect to the order of dependency, this type of solutions can hardly capture high-order dependence. 

% \textbf{add some discussion about concrete models, like PMFC. Emphasize the long-term dependence is always the bottleneck.}

%However, by nature, the long-range actions have long distance towards the target action and the volume of long-range actions accumulated in between is large, such that the structure of long-term dependence intrinsically are difficult to recognize. Markov models with simplified assumptions can hardly accurately recognize the structure of the long-term dependence, therefore failing in capturing the dependence accurately. 
% Hawkes process based methods are explored to capture the long-term dependence for sequential recommendation~\cite{cai2018modeling, wang2019modeling, du2015time}. ~\citet{du2015time} proposed a novel framework connecting Hawkes process with low-rank model to capture the dependence among actions. ~\citet{cai2018modeling} proposed a mixture of Hawkes process to capture the long-term dependence and the short-term dependence, which is composed of a multi-dimension Hawkes process and a uni-dimension Hawkes process. These models assume that historical actions have linear accumulative effect on next action. But the dependence among actions may be non-linear and even much more complex, which these models would fail in capturing. 

To address the limitations in Markov models, RNN and its its variants, Gated Recurrent Unit (GRU)~\cite{chung2014empirical} and Long Short-Term Memory (LSTM)~\cite{hochreiter1997long} based neural sequence models have been introduced to sequential recommendation \cite{hidasi2015session, wang2018neural, wang2019recurrent, li2017neural, hochreiter1997long}. \citet{hidasi2015session} applied RNNs to model users' behaviors in sessions; and their results show that RNN can effectively model sparse sequential data. \citet{li2017neural} used attention mechanism to emphasize influence from important past actions in the same session, which are expected to capture dependence among actions more accurately. \citet{guo2019streaming} proposed a  Streaming Session based Recommendation Machine (SSRM) to address the problem of session recommendation in streaming environment. Although these models are believed to be more powerful in modeling dependency among sequential actions than Markov models, they limit the scope to in-session observations. The modeling of longer term dependency, e.g., actions across sessions, is still lacking. As the long-term dependence is accumulated over a larger number of historical actions, its structure is expected to be complex. 

Moreover, inspired by the success of the self-attention architecture in the tasks of sequence to sequence learning, such as machine translations~\cite{vaswani2017attention, bahdanau2014neural}, researchers have recently explored self-attention in sequential recommendations, such as \cite{kang2018self,sun2019bert4rec}. Though these models achieved some encouraging results on sequential recommendation, the complex structure of the long-term dependence is not touched. 

Some attempts have been made recently to model long-term dependence for sequential recommendation. \citet{tang2019towards} proposed a mixture of neural models, i.e., Multi-temporal-range Mixture Model (M3R), by feeding the entire sequence into a self-attention model. But it does not explicit consider the structure of long-term dependence.  \citet{quadrana2017personalizing} proposed a hierarchical recurrent neural network model to jointly capture the transition among actions within and across sessions. The long-term dependency is thus structured in accordance to the session boundary. As a follow up work, \citet{you2019hierarchical} proposed a similar hierarchical model by replacing within-session RNN with a temporal convolutional network to speed up model training. But as we discussed before, structuring the dependency by session boundary forces actions in the same session to share the same influence from historical actions. It thus cannot provide accurate input for sequential recommendation. To address this limitation, we contextualize the structure of the long-term dependence based on inferred user intent, i.e., actions to be predicted can now condition on different subset of past actions. And we infer user intent by predicting the category of associated items.

\section{Method}
In this section, we introduce our proposed solution GatedLongRec. Without loss of generality, we assume there are a set of users $u\in U$, a set of items $v\in V$ and a set of categories $c\in C$ of the items. We denote an action as a tuple $a_n=(v_n, c_n)$, where $n$ is the index of the action in a sequence, $v_n$ is the item that the user has interacted with and $c_n$ is the category of the item. Different actions may associate with the same item. A sequence of $N$ actions generated by a user $u$ is denoted as $S_u=\{a_1, a_2, ..., a_N\}$, which is listed in the chronological order with respect to the timestamp of actions. In addition, the sub-sequence composed of a series of actions of the same category $c$ in the sequence $S_u$ is denoted as $S^c_u=\{a^c_1, ..., a^c_L\}$, where $a^c_1$ denotes the first action in the sub-sequence $S^c_u$, but it is not necessarily the first action in the sequence $S_u$. Given historical $N$ actions, the goal of sequential recommendation is to rank a subset of items for the current action $a_{N+1}=(v_{N+1}, c_{N+1})$ to satisfy the user's interest at this moment.

%We first describe our notations and problem setup. Then, we describe the overview of our model, followed by the detailed discussions of each component in our model. 

\begin{figure*}[t]
  \includegraphics[width=\textwidth]{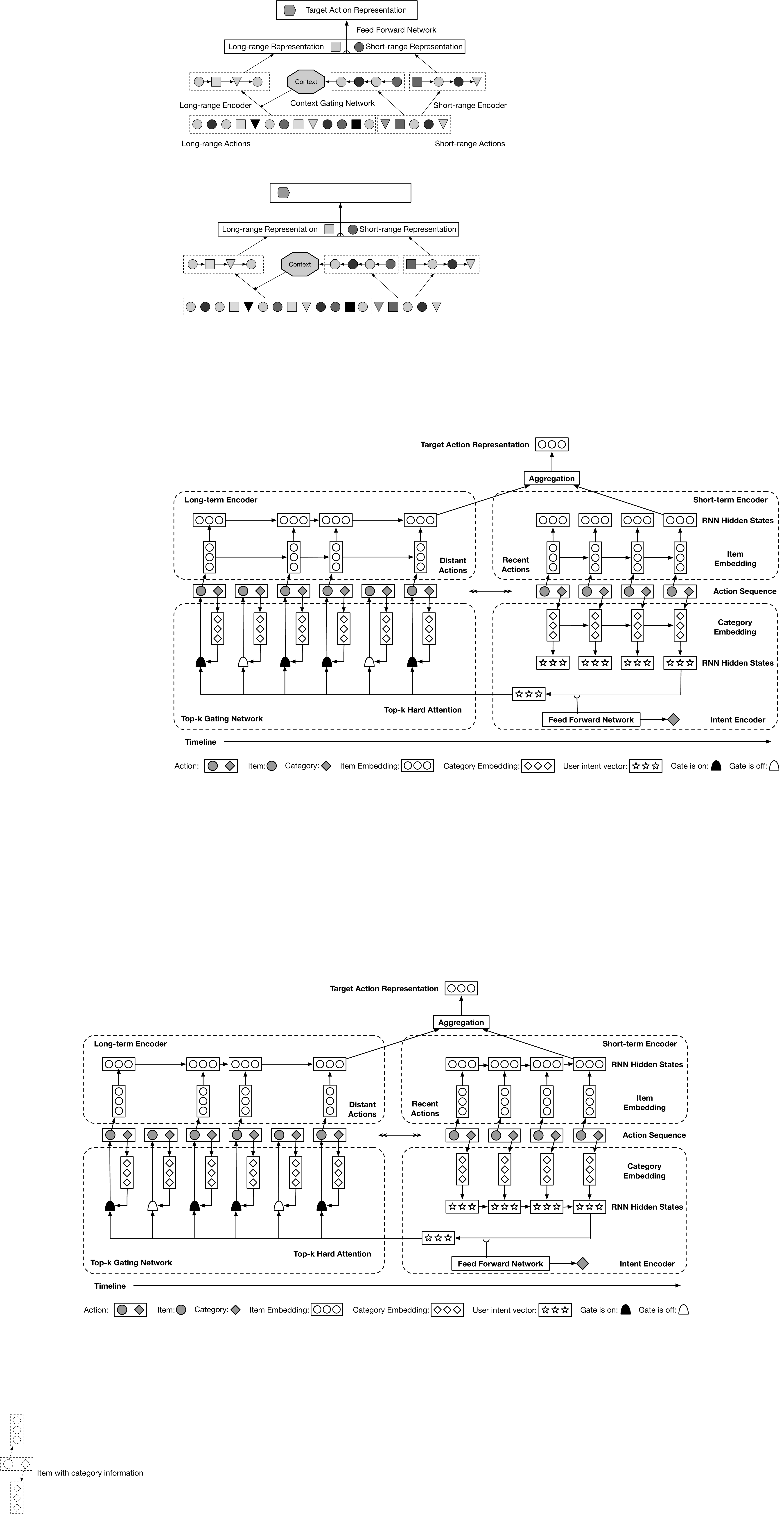}
  \caption{An overview of the proposed GatedLongRec model. To structure the long-term dependence, the intent encoder learns a vector representation of the user's intent, and the top-$k$ gating network identify distant actions from top-$k$ related categories to the user's ongoing intent for the modeling of long-term dependence. The long-term encoder takes these actions as input to model the long-term dependence. In the meanwhile, the short-term dependence is captured by the short-term encoder. These two types of dependence are intergated to predict the next action.} 
  \label{fig:GatedLongRec_overview}
\end{figure*}

%\subsection{Notations and Problem Setup}
 %To measure the quality of the recommendations, we evaluate whether the ground truth item $v_{N+1}$ appears in the recommended list.

% to predict the item $v_{N+1}$ that the user would interact with in the next action $a_{N+1}$ (target action) and to recommend a subset of items to meet users' need. The prediction of the item is denoted as $\hat{v}_{N+1}$.

\subsection{GatedLongRec}
The user's interest at action $a_{N+1}$ is influenced by past actions, including both recent and distant actions. For example, as illustrated in Figure \ref{fig:Sequence_example_intro}, a user's choice of which cellphone to examine is affected by her/his past browsing history of items in this category, e.g., revisit a previously examined one or exploring a new one. With respect to the sequential order of past actions, we define actions which are less than $M$ steps towards to the action to be predicted, i.e., $a_{N+1}$, as recent actions and those are are more than $M$ steps away as distant actions. $M$ is considered as a hyper-parameter, we will empirically study its impact in our experiments. Because of users' extensive interactions with an online service system, tens or hundreds of actions can be easily accumulated in the log data. The high volume and diversity in such historical actions introduce considerable complexity to modeling the dependence structure among actions.

To structure the long-term dependence at a fine granularity, we develop the GatedLongRec model, which infers users' ongoing intent to differentiate the influence from distant actions. GatedLongRec employs an intent encoder to obtain intent representation. Because of its close correlation to the user intent and ubiquitousness, category is utilized as a proxy for the intent representation learning. Based on the inferred user intent, GatedLongRec employs a gating network to identify a subset of distant actions for modeling the long-term dependence on each specific action to be predicted. Specifically, the gating network attends distant actions of a particular category with hard attention scores. Due to the uncertainty in the inferred user intent, top-$k$ categories are taken into consideration for long-term dependence modeling. In each specified category, GatedLongRec uses a shared long-term encoder to create a summary vector for the corresponding distant actions. Consequently, different subsets of past actions are considered to have different long-term influence on the current action, with respect to current user's intent. 

The general schematic description of GatedLongRec is shown in Figure~\ref{fig:GatedLongRec_overview}. We will introduce each component in a bottom-up manner, beginning with the input and progressing to the output. 

\textbf{Intent Encoder.}
% why encode intent?
The underlying intent drives users to interact with an online service system, and it reveals the structure of long-term dependence. With a particular intent behind the current action, like revisiting a previously examined cellphone, the user's choice of an item would be influenced by previously examined ones. Hence, inferring the intent driving the current action helps to realize its dependence on the distant actions. 

Since the user intent is latent, we cannot directly use it as input. As \citet{zhu2018learning} find that the category information of a user's interacted items closely correlates with his/her intent, we treat the category of the current action as a proxy of the user intent. Specifically, we treat the category as the supervision of learning the representation of the user intent reflected on the category. On the other hand, because of their short distance to the current action, features from the recent actions can be extracted to predict the intent behind the current action. To leverage this, GatedLongRec employs a RNN-based intent encoder to construct a representation of the current intent from recent actions, which should predict the category of corresponding action. 
 
%  believe a good intent representation should precisely predict the category.

Specifically, the intent encoder takes categories $\{c_{N-M+1}, ..., c_N\}$ of the $M$ most recent actions $\{a_{N-M+1}, ..., a_N\}$ as input, which are denoted as the diamonds in the Figure~\ref{fig:GatedLongRec_overview}, and map them into dense vectors $\{e^{cate}_{N-M+1}, ..., e^{cate}_N\}$ through an embedding layer $E_{cate} \in \mathbb{R}^{d_c \times |C|}$ of categories. Then we run a RNN over these dense embedding vectors to obtain the hidden representation for the user's intent of the action $a_{N+1}$. For the choice of RNN, either vallina RNN or other variants LSTM or GRU should satisfy the need. In our empirical evaluations, GRU obtained the best performance; and we adopt GRU in our discussion in this section. The update of the hidden state $h_i$ at step $i$ in GRU is as follows:
\begin{align*}
    r_i &= \sigma(W_r[h_{i-1}, e_i]), \\
    z_i &= \sigma(W_z[h_{i-1}, e_i]), \\
    \Tilde{h}_i &= tanh(W_{\Tilde{h}}[r_i \odot h_{i-1}, e_i]), \\
    h_i &= z_i\odot h_{i-1}+(1-z_i) \odot \Tilde{h}_i
\end{align*}
where $W_r, W_z$ and $W_{\Tilde{h}}$ are parameters to be learnt, $\odot$ represents elementwise product and $\sigma$ is the sigmoid function. We use the hidden state $h^{cate}_N$ at step $N$ as the intent representation, which is obtained via
\begin{equation*}
    h^{cate}_N = GRU (h^{cate}_{N-1}, e^{cate}_{N}), ~~ h^{cate}_N\in \mathbb{R}^{d^{cate}_h \times 1}
\end{equation*}
To predict the category $\hat{c}_{N+1}$, we feed the intent representation into a feed-forward layer and then a softmax layer. To reduce the complexity of GatedLongRec, the feed-forward layer has the same set of parameters as the embedding layer for the category, i.e., $d^{cate}_h = d_c$. The distribution of predicted categories is obtained via
\begin{equation*}
    p(\hat{c}_{N+1}=c)  = \frac{\exp(sim_{c})}{\sum_{c'}^{|C|} \exp(sim_{c'})},
    ~\text{where} \: sim_{c} = h^{cate}_N \cdot e^{cate}_{c} 
\end{equation*}
% We use cross entropy as the loss function of the category prediction, as
% \begin{align}
%     loss_{cate} = -\sum_{c}^{C} \delta(c_{N+1}=c) \log  p(\hat{c}_{N+1}=c).
% \label{eq:loss_cate}
% \end{align}

As the Figure~\ref{fig:GatedLongRec_overview} shows, the intent representation $h^{cate}_N$ is then fed into the gating network to identify related distant actions. We should notice that because the supervision of the category prediction has shorter path to the intent representation than the supervision of item prediction, the updating gradient of the user intent representation would be mainly determined by the supervision of category prediction~\cite{luong2015effective}. Thus, the learnt user intent representation mainly reflects the category-level user intent, and its uncertainty towards the user intent on the item-level exists.
% How to encode intent?
% What encoded?

\textbf{Top-$k$ Gating Network.} As a user's interest evolves over time, the large volume of accumulated actions are driven by a diverse set of intent. The distant actions driven by the intent irrelevant to the current action should have little or even no influence on its choice of item. As a result, the inferred user intent provides guidance on which subset of distant actions should be considered for the long-term dependence. 

To contextualize the long-term dependence regarding the user's intent at the current action, GatedLongRec employs a gating network to attend distant actions through hard attention. Attention model is widely used in sequence to sequence learning~\cite{vaswani2017attention, luong2015effective}, including soft attention and hard attention. Soft attention, adopted by self-attention models, weights each input with the corresponding attention score~\cite{bahdanau2014neural}. In contrast, hard attention attends one input, which puts sharp weight on one input~\cite{xu2015show}. Because of the large size of the distant actions, important actions should be clearly emphasized so that the long-term dependence can be recognized. Moreover, uncertainty exists in the inferred user intent, which is derived from the observed category information. Based on hard attention we propose a top-$k$ gating network which attends top-$k$ categories, instead of only the top-$1$ category. Furthermore, the top-$k$ gating network merges $k$ recommendation lists with attention weight to reflect the uncertainty of the user intent, which we describe in the later subsections. 

The gating network takes the intent representation $h^{cate}_N$ as the key and the embeddings of categories of distant actions as queries to compute the attention score, via a dot product. The attention weight $\alpha_c$ of the category $c$ is computed by  
\begin{align*}
    \alpha_c = \operatorname{softmax}\left( \boldsymbol{h^{cate}_N} \cdot \boldsymbol{e^{cate}_{c}}^T \right) 
\end{align*}
The weight indicates the probability of the user intents to interact with the items of category $c$. Distant actions belonging to categories with the top-$k$ attention scores  are extracted as the input to the long-term dependence encoder. In this way, distant actions which are driven by the ongoing intent are leveraged for modeling long-term dependence. As the Figure~\ref{fig:GatedLongRec_overview} shows, the gates of extracted distant actions turn on. With the change of the user intent representation $h^{cate}_N$, the extracted actions also vary. Consequently, the long-term dependence is contextualized based on the inferred user intent.  

% sim_{c} & = \operatorname{softmax}(h^{cate}_N \cdot E_{cate})

\textbf{Long-term Encoder.}
The transition patterns among the identified distant actions shed light on the items that the user would be interested in at the current moment. For example, among a series of viewed cellphones over time, when the user shifts from brand A to brand B, it strongly indicates the new focus on this user. Because the transition pattern is generally high-order and non-linear, we utilize a neural encoder to leverage the information buried in the identified distant actions. As RNN or its variants, i.e., LSTM and GRU are expressive to learn representations that encode high-order and non-linear patterns, GatedLongRec employs a RNN based long-term encoder. As we choose GRU for its promising empricial performance. 

For each category among the top-$k$ categories, the long-term encoder learns a vector encoding the transition patterns across actions of the category, thus resulting in $k$ vectors summarizing long-term dependence. Specifically, for actions $\{a^c_1, ..., a^c_L\}$ of the category $c$, denoted as circle in the Figure~\ref{fig:GatedLongRec_overview}, we project items associated with these actions into dense vectors $\{e^c_{1}, ..., e^c_L\}, e^c_i\in \mathbb{R}^{d_e \times 1}$ via the embedding layer $E_{item}$. The GRU in the long-term encoder encodes these dense vectors into a $d_l^{item}$ dimensional hidden representation, 
\begin{align*}
    h^{c}_i = GRU (h^{c}_{i-1}, e^c_{i}), \: i=1, ..., L
\end{align*}
In this way, $k$ vectors $\{h^{c}_L\}$ summarizing transition patterns of top-$k$ categories are obtained and integrated with the short-term dependence for item prediction. 

Because GatedLongRec extracts actions concerning the user's intent, the transition patterns among actions of the same category become evident to recognize. Without contextualizing the long-term dependence, like HRNN or HierTCN, the large volume of actions of other categories may make the dependence unidentifiable. Take Figure \ref{fig:Sequence_example_intro} for example,  the interleaved actions of examining handbags in between the examination of cellphones would dilute the model's attention on modeling users' preference in both cellphones and handbags. By separating the past actions by the categories of the user's current interest, the model can recognize user preference and therefore provide relevant recommendations.  

% \textbf{add some discussion about the transition pattern captured}

\textbf{Short-term Encoder.} Apart from the deliberately structured long-term dependence, the short-term dependence is also necessary for the model to accurately understand the user interest. Because of the shorter distance, recent actions generally leave the user with stronger impressions~\cite{cai2018modeling}, thus influencing his/her choice of item at the current action. Use Figure \ref{fig:Sequence_example_intro} as an example again: when examining cellphones, by recalling the attributes of last examined cellphones, the user can make better decisions on what to examine next. This close correlation is also empirically verified by previous work~\cite{}. Because many models have been proposed for this type of dependence, we do not deliberately model it in the work and adopt the most common and well analyzed model, i.e., RNN, for the short-term dependence. Due to the better performance in experiments, GatedLongRec employs a GRU-based short-term encoder. It takes the most recent $M$ actions $\{a_{N-M+1}, ..., a_{N}\}$ as input and projects their associated items into dense vectors $\{e_{N-M+1}, ..., e_N\}, e_i\in \mathbb{R}^{d_e \times 1}$ via the embedding layer of items. The GRU in the encoder encodes these embeddings into a $d^{item}_s$ dimensional hidden vector via
\begin{align*}
    h^{short}_{i}= GRU(h^{short}_{i-1}, e_i), \: i=N-M+1, .., N
\end{align*}
The representation $h^{short}_{N}$ captures the transition pattern among recent actions, which is combined with long-term dependence for item prediction.

\begin{table*}[]
\centering
\caption{Statistics of two evaluation datasets.}
\label{table:Statistics}
\begin{tabular}{@{}cccccccc@{}}
\toprule
\textbf{Dataset}   & 
\textbf{\#Users} &
\textbf{\#Items}  &
\textbf{\#Categories}  &
\textbf{\begin{tabular}[c]{@{}c@{}}\#Avg actions \\ per user\end{tabular}} & \textbf{\begin{tabular}[c]{@{}c@{}}\#Train actions\end{tabular}} &
\textbf{\begin{tabular}[c]{@{}c@{}}\#Validation actions\end{tabular}} &
\textbf{\begin{tabular}[c]{@{}c@{}}\#Test actions\end{tabular}}  
\\ \midrule
\textbf{Taobao} &  51,275   &  68,007   &   201  &  73.836  & 2,995,258  & 430,797  & 359,906 \\
\textbf{News}  & 54,603 &  178,643  & 58  & N/A  &   N/A &   N/A  & N/A \\ \bottomrule
\end{tabular}
\end{table*}

\textbf{Item Prediction.}
To utilize both the short-term dependence and the long-term dependence for sequential recommendation, we fuse their corresponding representations to predict the user's interest of the current action. To satisfy the user's information need, we match the user's interest with the most similar items as our recommendation.

With the output of the short-term encoder $h^{short}_N$ and the output of the long-term encoder $\{h^{c}_L\}, c=\text{top-}1, .., \text{top-}k$, GatedLongRec concatenates $h^{short}_N$ with these $\{h^{c}_L\}$ separately as 
\begin{align*}
  h^c=W_{f}[h^{c}_L, h^{short}_N]
\end{align*}
Thus $k$ hidden vectors $h^c\in \mathbb{R}^{d_f}$ representing the user interest at the target action. We should notice that as these vectors encode the transition patterns among items, different from the category information encoded by the proxy user intent representation, they are utilized for the recommendations. GatedLongRec feeds these $k$ user vectors individually into a feed-forward layer which has the same parameters as the embedding layer of items, and then individually into a softmax layer to produce $k$ separated distributions of items. To handle the potentially large number of recommendation candidates in practice, we apply negative sampling to compute the loss by drawing $Z$ items as negative samples according to their popularity. Therefore, when the user intents to interact with items from category $c$, the probability of the item $v$ appearing in the recommendation for action $a_{N+1}$ is
\begin{equation}
    p(\hat{v}_{N+1}=v|c) = \frac{\exp(sim^{{c}}_{v})}{\sum_{v'}^{Z+1} \exp(sim^{{c}}_{v'})}, 
    ~\text{where}\:  sim^{{c}}_{v} =  {h^{{c}}}^\top E_{item}.
\label{eq:item_pred}
\end{equation} 
These $k$ distributions correspond to the top-$k$ most likely user intent inferred by the gating network. To account for the uncertainty of the inferred user intent, the $k$ distributions of items are combined by the attention weight $\alpha$ obtained in the top-$k$ gating network into a distribution of items for recommendation, as 
\begin{align}
    p(\hat{v}_{N+1}=v) = \sum_{{c}\in \{\text{top-}1, ..., \text{top-}k\}} p(\hat{v}_{N+1}=v|{c})\alpha_{{c}}.
    \label{eq:item_joint_prob}
\end{align}

% To estimate the parameters needed for item prediction, we adopt the cross entropy as the loss function, as 
% \begin{align}
%     loss_{item} = -\sum_{v=1}^{Z+1} \delta(v_{N+1}\!=\!v)\log p(\hat{v}_{N+1}=v).
% \label{eq:item_loss}
% \end{align}

\subsection{Network Training}
GatedLongRec outputs both the item and its category that the user would be interested in for the next action $a_{N+1}$. During training, both the observed item and its category of the action $a_{N+1}$ in the ground truth data serve as supervision. 

\noindent\textbf{Loss function.} 
To estimate parameters of GatedlongRec, we use cross entropy as the loss function of the category prediction, as
\begin{align}
    loss_{cate} = -\sum_{c}^{C} \delta(c_{N+1}=c) \log  p(\hat{c}_{N+1}=c).
\label{eq:loss_cate}
\end{align}

In addition, we also adopt the cross entropy as the loss function of item prediction, as 
\begin{align}
    loss_{item} = -\sum_{v=1}^{Z+1} \delta(v_{N+1}\!=\!v)\log p(\hat{v}_{N+1}=v).
\label{eq:item_loss}
\end{align}

GateLongRec jointly optimizes the item prediction and the category prediction. Therefore, the joint loss is  
\begin{align*}
    loss = \lambda\times loss_{item} + (1-\lambda)\times loss_{cate}.
\end{align*}
where $\lambda$ is the hyper-parameter controlling the weight of these two losses in the objective function. Unlike training, the category of the action to be predicted is unknown during inference. This difference of category distribution may hurt the performance during inference~\cite{zhang2019bridging}. To mitigate this issue, we modify the training schedule to enable the model to adopt to the condition. At the early stage of the training, we utilize the joint loss from both item prediction and category prediction for parameter estimation; and at the late stage of the training, we utilize only the loss of item prediction to update the model. This means we change the hyper-parameter $\lambda$ from $0.5$ to $1$.

During training, we estimate the parameters of GatedLongRec. During inference, GateLongRec computes probabilities of items according to Eq.~\eqref{eq:item_joint_prob}, ranks items according to these probabilities and recommends top-ranked items to the user.

\section{Experiments}
In this section, we study the effectiveness of our model on providing sequential recommendations. We first describe two large datasets used for evaluation, followed by the implementation details of our model on these two datasets. To show that structuring the long-term dependence improves the quality of recommendations, we compare our model against a comprehensive set of baselines. Moreover, we perform ablation analysis to understand the importance of the top-$k$ gating network, the intent encoder and the long-term encoder to the modeling of the long-term dependence through experiments with different hyper-parameters.  

\subsection{Datasets}
We use a public Taobao dataset~\footnote{https://tianchi.aliyun.com/dataset/dataDetail?dataId=649} and a proprietary news recommendation dataset for evaluation. The public Taobao dataset contains sequences of users' actions on an online shopping website~\footnote{https://www.taobao.com/}. Each action is associated with the user ID, item ID, item's category ID, action's timestamp and behavior type (click, add to cart. etc.). Due to privacy concern, the semantic meaning of the categories is not disclosed. We randomly sampled 100,000 users' sequences from the dataset for  evaluation. We used observations from November 25, 2017 to December 3, 2017, where actions in the first 7 days were used as training set, actions in the 8th day as validation set, and actions in the 9th day as test set. We removed items associated with less than 20 actions, and removed users with less than 20 or more than 300 actions. We merged categories which have fewer than 100 items into a special category, represented as category ``UNK''. 

For the news recommendation dataset, we collected logs of randomly sampled 100,000 users from a news media website. The time span of the log is from April 2, 2019 to April 15, 2019. We use actions in the first 13 days as the training set, actions from April 15 12:00am to April 15 12:00pm as the validation set, and actions from April 15 12:00pm to April 15 11:59pm as the test set. We removed items with less than 20 actions, and removed users with less than 50 or more than 1000 actions. We merged categories with fewer than 500 items into the ``UNK'' category. The basic statistics of our evaluation datasets is shown in Table~\ref{table:Statistics} ~\footnote{Subject to the website's business policy, several fields of News dataset are filled in with N/A.}. 

% We randomly sample 100,000 users from the public Taobao dataset. The dataset contains user ID, item ID, item's category, item's timestamp and behavior type (like ). We keep actions of different behavior types from November 25, 2017 to December 3, 2017. We use actions in the first 7 days as train dataset, actions in the 8th day as validation dataset, and actions in the 9th day as test dataset. We remove items associated with less than 20 actions, and remove users with less than 20 or more than 300 actions. We merge categories which have fewer than 100 items. On the other hand, we build the news dataset by collecting logs of randomly sampled 100,000 users in two weeks from April 2, 2019 to April 15, 2019. We use actions in the first 13 days as train dataset, actions from April 15 12:00am to April 15 12:00pm as validation dataset, and actions from April 15 12:00pm to April 15 11:59pm as test dataset. We remove items with less than 20 actions, and remove users with less than 50 or more than 1000 actions. We merge categories which have fewer than 1000 items. The data statistics is as the Table~\ref{table:Statistics} shows.

\subsection{Experiment Setup}
\textbf{Implementation Details.} On Taobao dataset, we keep the items' embedding in the input side to be the same as that in the output side in our model, and the dimension $d_e$ is chosen from $\{128, 200, 300, 400\}$. As we found that $d_e=300$ led to better results, the results of our model on Taobao dataset were obtained with $d_e=300$. The dimension $d_s^{item}$ of the hidden representation learnt by short-term encoder is chosen from $\{128, 200, 300\}$ and we set $d_s=300$ because of better results. Likewise, we set the dimension $d_l^{item}$ of the hidden representation learnt by long-term encoder to 300. We kept the dimension $d_f$ of the hidden vector representing user's dynamic interest to be the same as $d_e$. Besides, the same category embedding layer $E_{cate}$ was used in both the input and output side of our model. The dimension $d_c$ of category embedding was set to 64. The size $d_h^{cate}$ of the hidden representation in the top-$k$ gating network was set to 64. We found that our model is sensitive to the learning rate and the optimal learning rate $0.001$ is chosen from $\{0.0001, 0.0005, 0.001, 0.01\}$. 

On the News dataset, the details of our model are that $d_e=400, d_s^{item}=400, d_l^{item}=400, d_f=400, d_c=64, d_h^{cate}=64$. The learning rate was set to $0.0005$. For both datasets, the dropout rate of GRU was set to $0.2$. We used Adam as the optimizer \cite{kingma2014adam} and we change $\lambda=0.5$ to $\lambda=1.0$ when the loss of category prediction in validation set does not decrease. We stop training until the joint loss in validation set does not decrease for 10 consecutive epochs. When investigating the influence of a particular hyper-parameter, we keep other hyper-parameters constant.

\begin{table}[ht]
\caption{Performance of models on two datasets.}
\vspace{-2mm}
\begin{center}
\begin{tabular}{>{\centering\arraybackslash}p{1.1cm}>{\centering\arraybackslash}p{3.4cm}>{\centering\arraybackslash}p{2.8cm}}
    \hline
    Methods&\begin{tabular}[c]{@{}cc@{}}\multicolumn{2}{c}{Taobao}\\ \hline
                        Recall@100 & MRR@100
        \end{tabular} & \begin{tabular}[c]{@{}cc@{}}\multicolumn{2}{c}{News}\\ \hline
                        Recall@100 & MRR@100
        \end{tabular} \\
    \hline
    GlobalPop & \begin{tabular}[c]{@{}cc@{}} 0.0257 & 0.0023
        \end{tabular} & \begin{tabular}[c]{@{}cc@{}} 0.0090 & 0.0035
        \end{tabular} \\ 
    SeqPop & \begin{tabular}[c]{@{}cc@{}} 0.3004 & 0.0645
        \end{tabular} & \begin{tabular}[c]{@{}cc@{}} 0.0001 & 0.0001
        \end{tabular} \\ 
    FOT & \begin{tabular}[c]{@{}cc@{}} 0.1830 & 0.0661
        \end{tabular} & \begin{tabular}[c]{@{}cc@{}} 0.0466 & 0.0083
        \end{tabular} \\ \hline
    GRU4Rec & \begin{tabular}[c]{@{}cc@{}} 0.2967 & 0.0882
        \end{tabular} & \begin{tabular}[c]{@{}cc@{}} 0.1731 & 0.0179
        \end{tabular} \\ 
    SASRec & \begin{tabular}[c]{@{}cc@{}} 0.3123 & 0.0869
        \end{tabular} & \begin{tabular}[c]{@{}cc@{}} 0.1618 & 0.0057
        \end{tabular} \\ 
    HierRNN & \begin{tabular}[c]{@{}cc@{}} 0.2244 & 0.0503
        \end{tabular} & \begin{tabular}[c]{@{}cc@{}} 0.1011 & 0.0031
        \end{tabular} \\ 
    M3R & \begin{tabular}[c]{@{}cc@{}} 0.2754 & 0.0901
        \end{tabular} & \begin{tabular}[c]{@{}cc@{}} 0.1793 & 0.0207
        \end{tabular} \\ 
    RNN+MTL & \begin{tabular}[c]{@{}cc@{}} 0.3023 & 0.0875
        \end{tabular} & \begin{tabular}[c]{@{}cc@{}} 0.1989 & 0.0221
        \end{tabular} \\ 
    \hline
    GatedLongRec & \begin{tabular}[c]{@{}cc@{}} \textbf{0.3665} & \textbf{0.0942}
        \end{tabular} & \begin{tabular}[c]{@{}cc@{}} \textbf{0.2351} & \textbf{0.0251}
        \end{tabular} \\ 
    \hline
\end{tabular}
\end{center}
\label{tab:baselines_vs}
\vspace{-2mm}
\end{table}

\noindent\textbf{Baselines.} We compare GatedLongRec with an extensive set of baselines for sequential recommendations. \\
$\bullet$
\textbf{Global Popularity (GlobalPop).} It ranks items according to their popularity in the training set in a descending order. This is a simple and strong baseline ~\cite{tang2018personalized, li2017neural}. But it ignores the dependency among actions in a sequence.
\\
$\bullet$
\textbf{Sequence Popularity (SeqPop).} It ranks items according to their popularity in the target user's sequence in a descending order. The popularity of an item is updated sequentially with more actions of the target user being observed. But it ignores the dependency among actions of various items in a sequence.
\\
$\bullet$
\textbf{First Order Markov Model (FOT).} This method assumes that each action depends only on the last action. It ranks items according to the probability of items given the item in the last action. And this transition probability is estimated from the training set. It only considers the first order dependency, and high-order dependence cannot be captured.
\\
$\bullet$
\textbf{Recurrent Neural Network (GRU4Rec).} ~\citet{hidasi2015session} proposed to use RNN (GRU) to model users' sequences of actions. But this model does not deliberately model the long-term dependence, but simply summarize it with a recurrent latent state vector. %The model is required to figure out the structure of the long-term dependence by itself.
\\
$\bullet$
\textbf{Self-attentive Sequential Recommendation (SASRec).}~\citet{kang2018self} proposed to use self-attention to model users' sequences of actions. It does not deliberately structure the long-term dependency. 
\\
$\bullet$
\textbf{Hierarchical Recurrent Neural Network (HierRNN).} ~\citet{quadrana2017personalizing} proposed a hierarchical RNN to model users' sequences of actions. The structure of the long-term dependence is hierarchical with respect to time-based sessions, which we have described in the related work section. 
\\
$\bullet$
\textbf{Multi-temporal-range Mixture Model (M3R).} ~\citet{tang2019towards} proposed a mixture of neural models to model users' sequences of actions. It separates the dependency among actions into tiny-range dependency, short-range dependency and long-range dependency by time. A self-attention model is used to capture the long-range dependence. 
\\
$\bullet$
\textbf{Category Based Recommender (RNN+MTL).} ~\citet{zhao2018categorical} incorporated the multi-task learning into sequential recommendation. They proposed a RNN-based model which predicts both the category and item of next action. This model also uses the category information of actions to improve recommendation. But it does not deliberately structure the long-term dependence. 

\noindent\textbf{Evaluation Metrics.}
To evaluate our model against other models in recommending items, we rank the ground-truth items against predictions by these models. The ground-truth is defined by items associated with the recorded actions. We use Recall@K and Mean Reciprocal Rank@K (MRR@K) as the performance metrics.
\\
$\bullet$
Recall@K: This metric counts the proportion of times when the ground-truth items are ranked among the top-K predictions by a model. 
% In this paper, we set K to be 100, as the number of candidate items in both datasets is huge.  
\\
$\bullet$
MRR@K: This metric reports the average of reciprocal ranks of the ground-truth relevant items. If the rank is larger than K, the reciprocal rank is 0.
%  We set K to be 100. 

\begin{figure*}[!h]
\centering
\begin{tabular}{c c c c}
\includegraphics[width=1.7in]{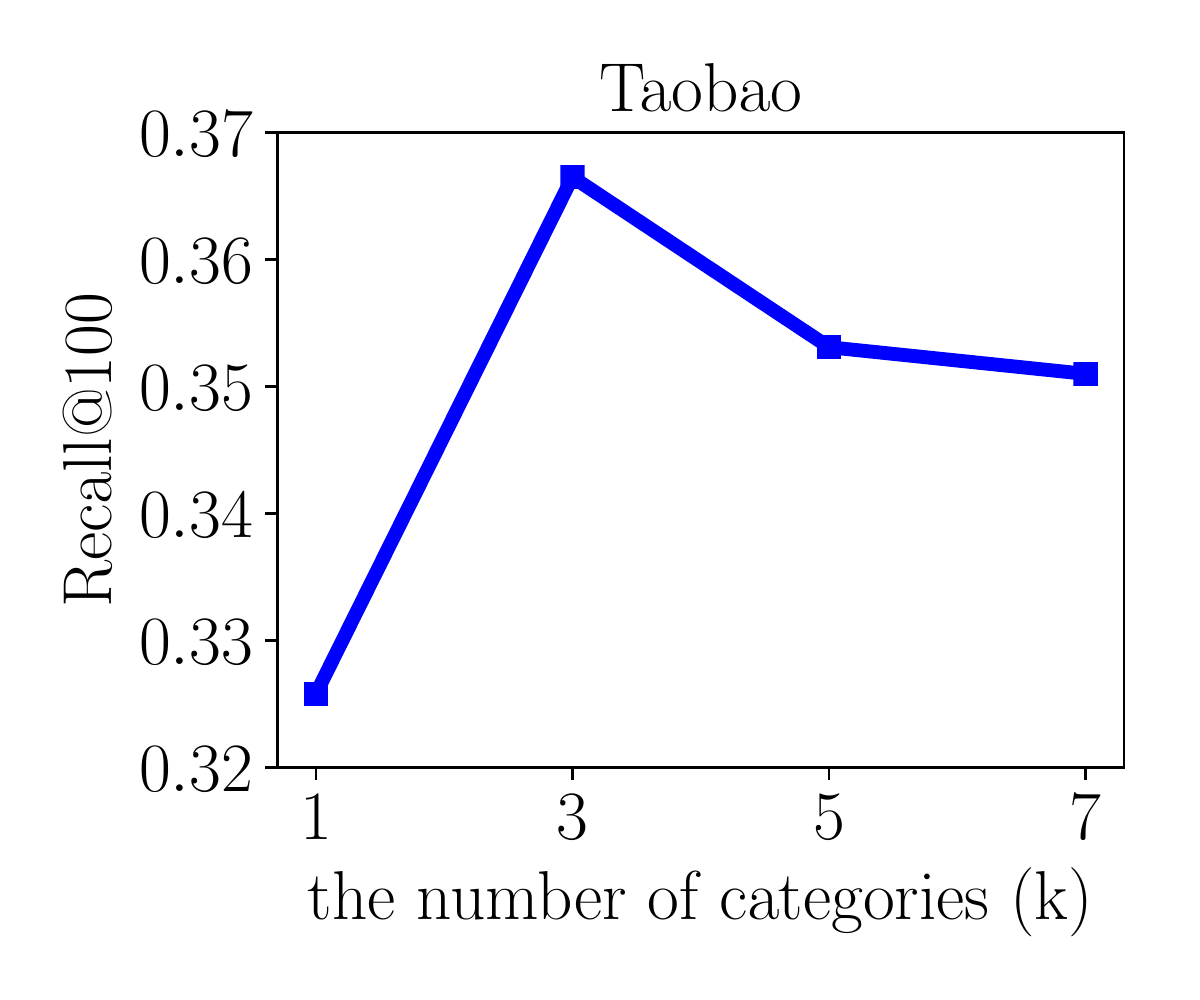} \hspace{-0.2 in} & \includegraphics[width=1.7in]{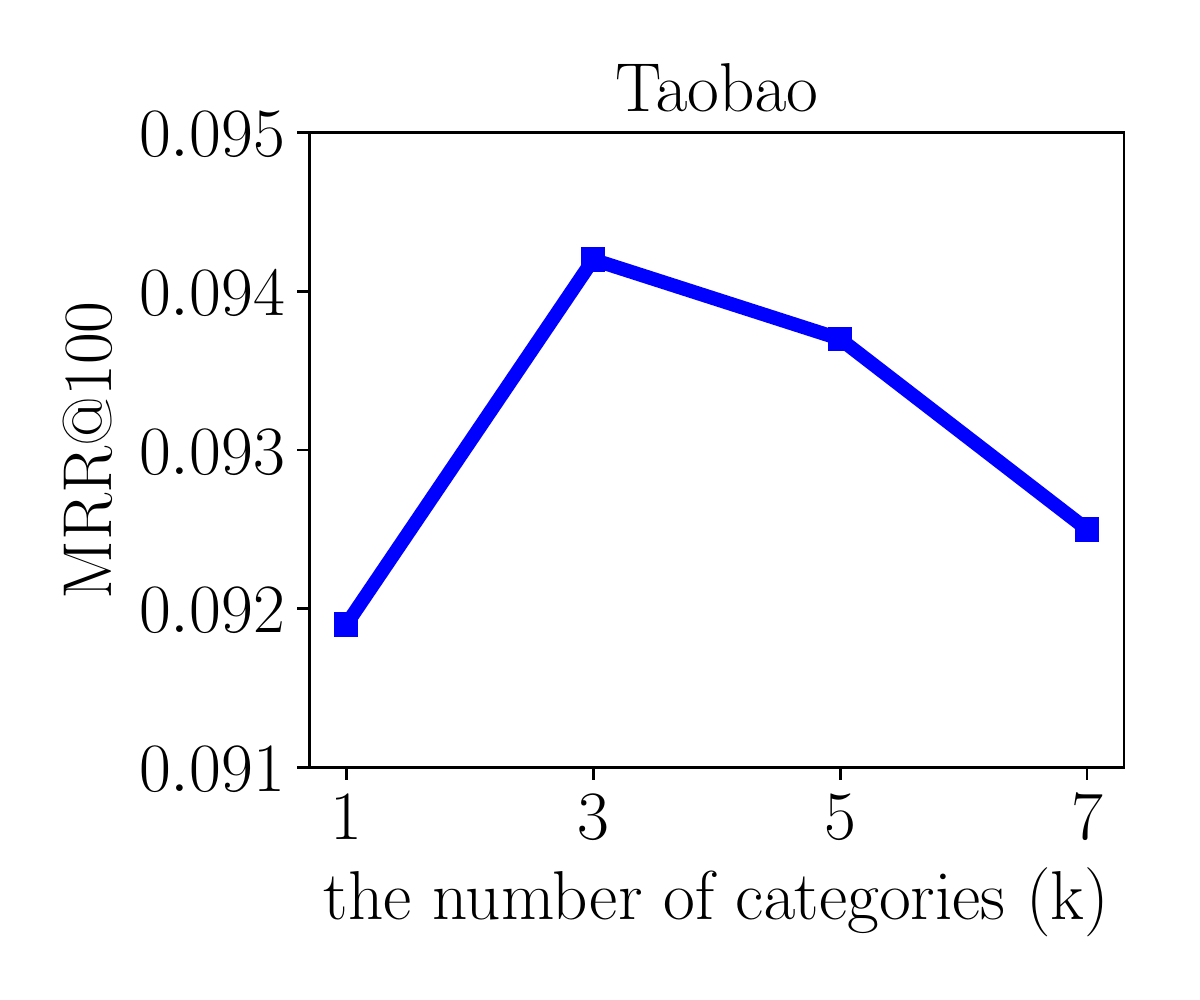} \hspace{-0.2 in} & \includegraphics[width=1.7in]{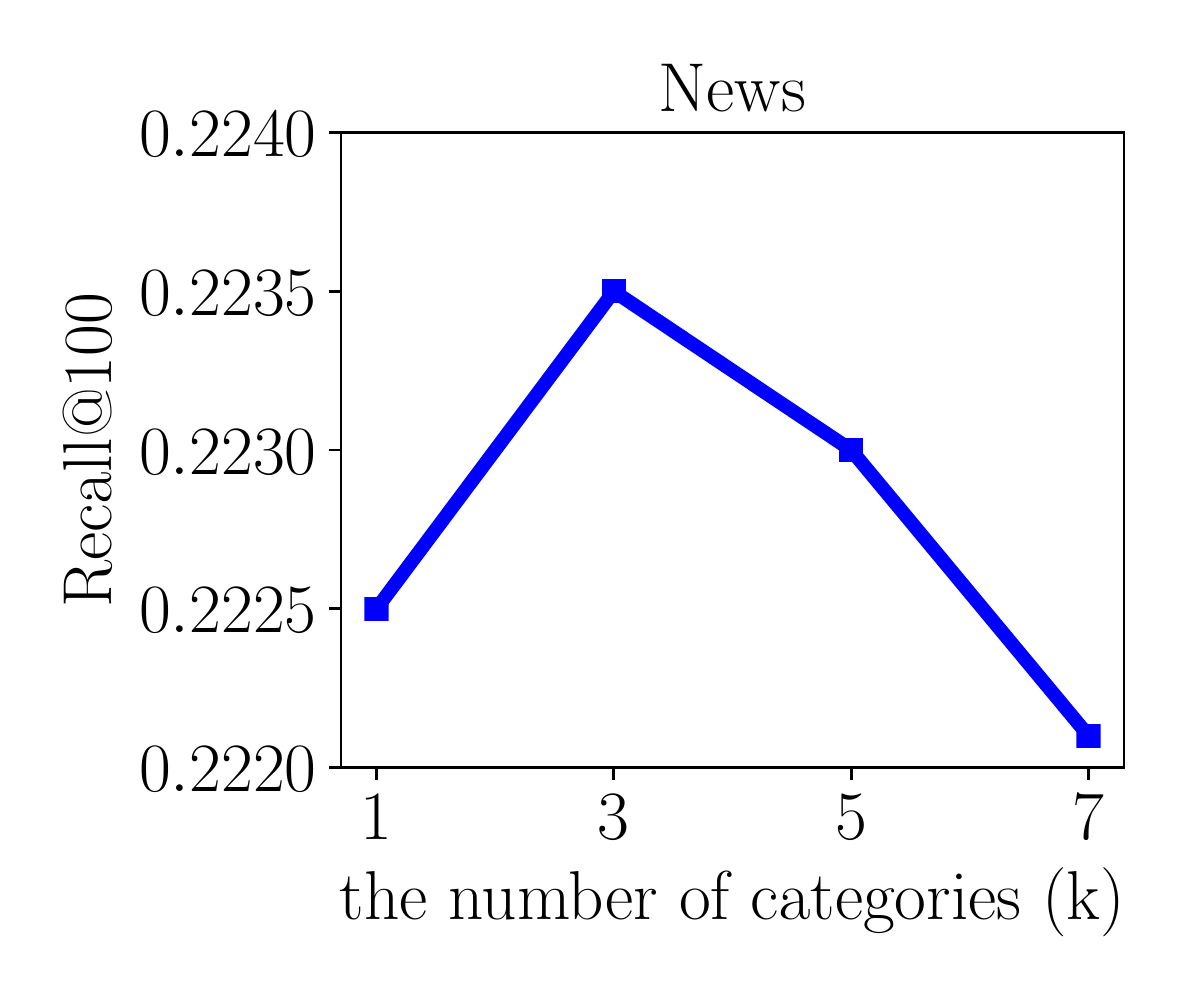} \hspace{-0.2 in} &
\includegraphics[width=1.7in]{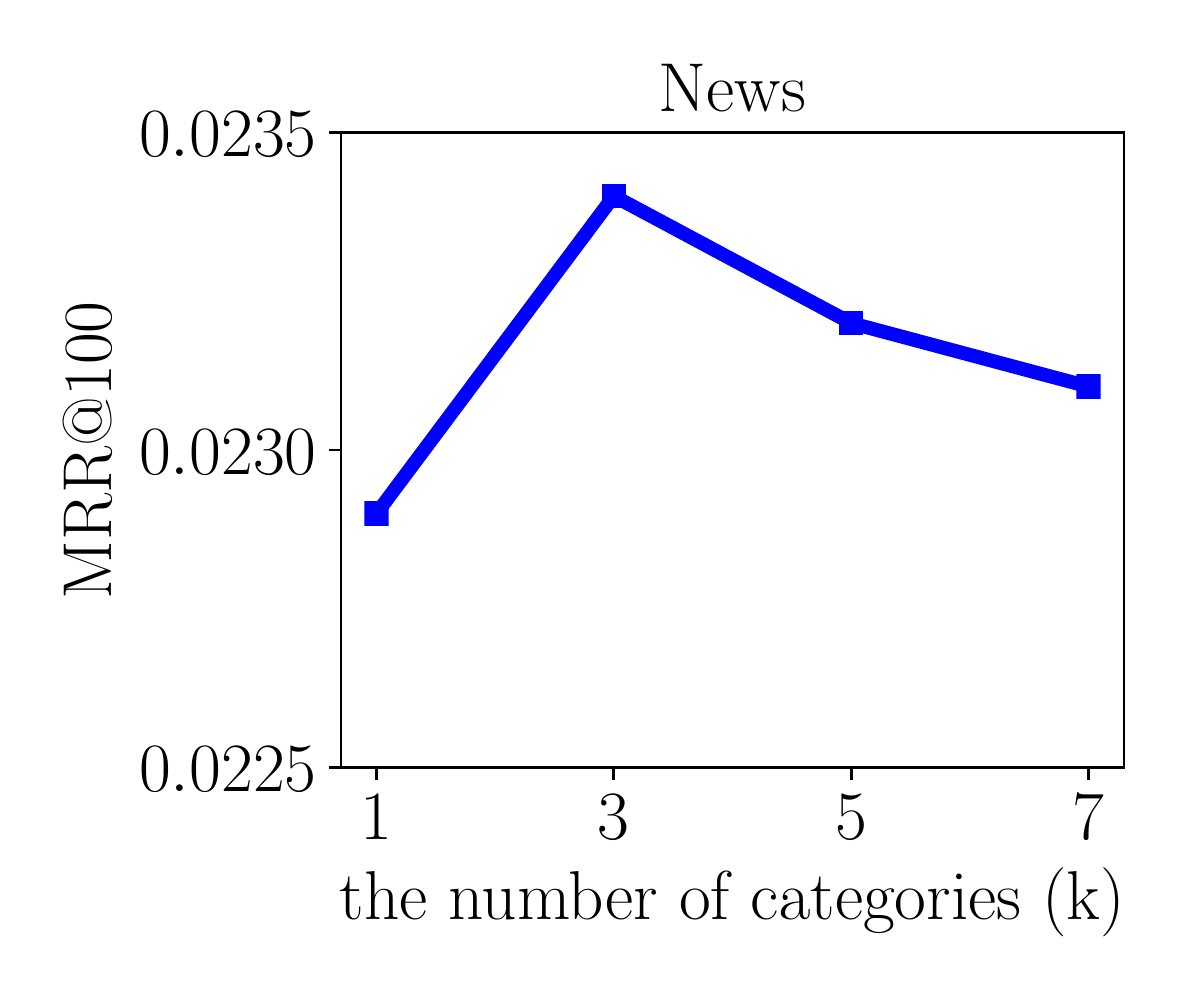} \\
\small{(a)} & \small{(b)} & \small{(c)} & \small{(d)}
% (a) tune $\beta_m$ with fixed $\beta_s$ & & (b) tune $\beta_s$ with fixed $\beta_m$
\end{tabular}
\caption{Performance of GatedLongRec with different number of categories ($k$) considered for the mixture of item predictions on two datasets.}
 \label{fig:GatedLongRec_mix}
\end{figure*}

\subsection{Comparison against Baselines}
On Taobao dataset, the setting of our model is that the most recent $M=20$ actions are considered as recent actions for short-term dependence. The rest are considered as distant actions. As the variance of the number of distant actions under each category is very large, for each category we take into account at most $T=20$ latest distant actions for the long-term dependence. The number of categories considered in the top-$k$ gating network is $k=3$. On News dataset, the setting is $M=20, T=20, k=3$. We analyze the influence of these hyper-parameters on the performance of our model in the next sections. 

The results of our model and baselines on two datasets are reported in the Table~\ref{tab:baselines_vs}. As the number of candidate items in both datasets is huge, we set K to 100 in our evaluation metrics. We find find that GatedLongRec outperforms all baselines on these two datasets. As users would repeat clicking an item at different times on Taobao, SeqPop achieves good performance on it. In contrast, as users are very unlikely to read a news article multiple times at different times, SeqPop performs poorly on News dataset. As SeqPop does not consider the dependence among actions of different items, it performs worse than GatedLongRec on both datasets. FOT considers the first order dependence among actions, but ignores high-order dependence. As a result, it underperforms GatedLongRec. 

Because both GRU4Rec and SAS4Rec take the whole sequence as input for the modeling of the dependence, the transition patterns among actions of a certain category, like cellphones in the Figure~\ref{fig:Sequence_example_intro}, can be diluted by the actions of other categories, like  handbags. Consequently, neither GRU4Rec nor SAS4Rec can recognize the user's preference at the target action accurately, thus performing worse than GatedLongRec. Although M3R separates the dependence into different types, using self-attention model for the long-term dependence prevents the model from recognizing the user's preference accurately. The same is observed in SAS4Rec. HRNN structures the long-term dependence hierarchically, but the hierarchical structure forces actions driven by different intents to share the same influence from past actions. As a result, the recommendations by HRNN become less discriminate, thus achieving worse performance than GatedLongRec. The comparison between RNN+MTL and GRU4Rec shows that category information generally helps the recommendation. This aligns with the observation that the categories are correlated with the underlying user intent, and therefore providing more helpful information to identify the next item that the user is interested in. The comparison of RNN+MTL with GatedLongRec demonstrates that using category as a proxy of the user intent and structuring the long-term dependence concerning the inferred user intent allows GatedLongRec to better utilize the category information, thus further improving the recommendation quality.

\subsection{Intent Encoder} 
GatedLongRec treats the category as the proxy for the user intent. As the category affects the learnt intent representation which then further affects the modeled long-term dependence, we investigate the correlation between the category and the intent representation by evaluating how accurately the  intent representation predicts the category. We compare using the learnt intent representation for prediction against other three methods, including ``GlobalPop\_cate'' using the global popularity of the categories for prediction, ``SeqPop\_cate'' using the popularity of the categories in the target sequence for prediction, and ``FOT\_cate'' using the first order transition probability of categories for prediction. The result is shown in the Table~\ref{tab:GatedLongRec_category}. We can observe that the learnt intent representation can outperform these models on predicting the category of the next action. This observation suggests that the representation learnt by the intent encoder well recognizes the user intent reflected on the categories, providing accurate signals to contextualize the long-term dependence.

Moreover, we could see that on Taobao dataset the category predictions are more accurate than those on News dataset. This difference may correlate with the characteristics of users' behaviors on these two online platforms. When shopping online, like on Taobao, a user's intent may be more consistent in a short-range than that when reading news, so that utilizing categories of recent actions to predict the category of the target action results in better performance on Taobao dataset. 

\begin{table}[ht]
\caption{Performance of category prediction with the proxy intent presentation on two datasets.}
\begin{center}
\begin{tabular}{ccc}
    \hline
    Models & \begin{tabular}[c]{@{}cc@{}}\multicolumn{2}{c}{Taobao}\\ \hline
                        Recall@3 & MRR@3
     \end{tabular} & \begin{tabular}[c]{@{}cc@{}}\multicolumn{2}{c}{News}\\ \hline
                        Recall@3 & MRR@3
        \end{tabular}\\
    \hline
    GatedLongRec &  \begin{tabular}[c]{@{}cc@{}} 0.7345 &  0.5516
        \end{tabular} &  \begin{tabular}[c]{@{}cc@{}} 0.5206 & 0.3213
        \end{tabular} \\
    GlobalPop\_cate & \begin{tabular}[c]{@{}cc@{}} 0.4482 & 0.2450
        \end{tabular} & \begin{tabular}[c]{@{}cc@{}} 0.4012 & 0.1895
        \end{tabular} \\
    SeqPop\_cate & \begin{tabular}[c]{@{}cc@{}} 0.6250 & 0.3890
        \end{tabular} & \begin{tabular}[c]{@{}cc@{}} 0.4721 & 0.2103
        \end{tabular} \\
    FOT\_cate & \begin{tabular}[c]{@{}cc@{}} 0.5709 & 0.2594
        \end{tabular} & \begin{tabular}[c]{@{}cc@{}} 0.4182 & 0.2047
        \end{tabular} \\
    \hline
\end{tabular}
\end{center}
\label{tab:GatedLongRec_category}
\end{table}

To look into how the learnt user intent representation affects the modeling of the long-term dependence, we decompose the performance of the item prediction with respect to the category prediction. We define the case when the ground-truth item ranks top-100 among item predictions as correct item prediction (corresponding to Recall@100). Otherwise it is defined as wrong item prediction. We also define the case when the ground-truth category ranks top-3 among the predicted ones as correct category prediction (corresponding to the top-3 gating network). Otherwise, it is defined as wrong category prediction. To study how item prediction performs given the category prediction, We visualize the ratio of correct item predictions under the correct and wrong category predictions. 

\begin{figure}[t]
\centering
\includegraphics[width=0.45\textwidth]{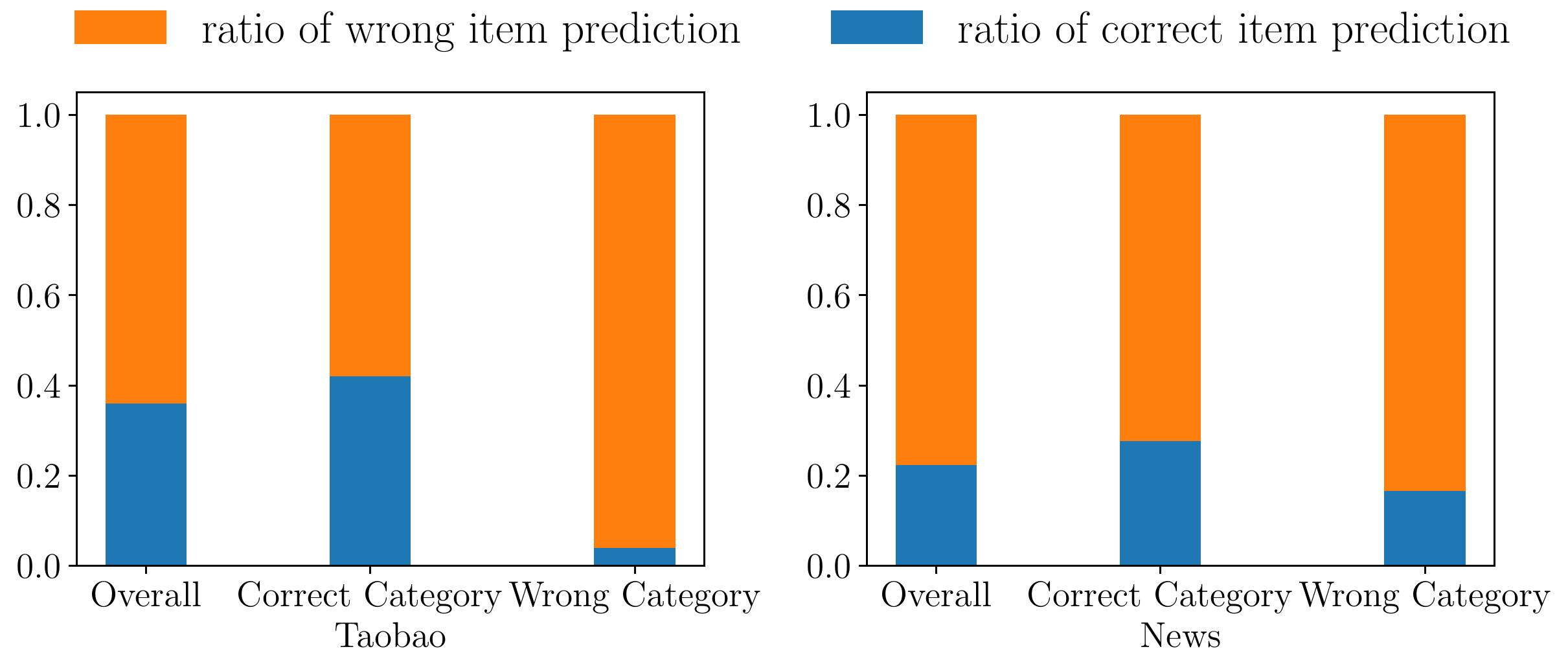} 
\caption{The overall ratio of the correct item predictions, the ratio of the correct item predictions given the correct category predictions, the ratio of the correct item predictions given the wrong category predictions. } 
\label{fig:GatedLongRec_cate}
\end{figure}

The result is as the Figure~\ref{fig:GatedLongRec_cate} shows. We can notice that when the user intent representation can predict categories correctly, the item predictions are much more accurate, compared with the situation when category predictions are wrong, especially on Taobao dataset. This observation suggests that learning user intent representation which can accurately predict category helps the modeling of the long-term dependence, thus improving the recommendation quality.

% the category prediction affects the item prediction in GatedLongRec, we decompose the performance of the item prediction with respect to the category prediction. We define the case when the ground-truth item ranks top-100 among item predictions as correct item prediction (corresponding to recall@100). Otherwise it is defined as wrong item prediction. We also define the case when the ground-truth category ranks top-5 among category predictions as correct category prediction (corresponding to 5 mixture predictions). Otherwise, it is defined as wrong category prediction. To see how item prediction performs given the category prediction, We visualize the ratio of correct item predictions under the correct and wrong category predictions, as the Figure~\ref{fig:GatedLongRec_cate} shows. We could see when category predictions are correct, the item predictions are much more accurate than those when category predictions are wrong, espeically on Taobao dataset. 

\subsection{Top-$k$ Gating Network}
Due to the uncertainty of the inferred user intent, we proposed a top-$k$ gating network to extract actions of top-$k$ categories with respect to the inferred user intent for the modeling of the long-term dependence. To investigate the importance of the top-$k$ gating network to GatedLongRec, we compare the performance of GatedLongRec with different values of $k$. The result is reported in Figure~\ref{fig:GatedLongRec_mix}.

We observe that on both datasets, when $k=1$, the performance of GatedLongRec is bad. When $k=1$, the top-$k$ gating network turns into hard attention, which  extracts actions of the most related category concerning the inferred user intent. Once the inferred user intent is inaccurate in regard to the true user intent driving the next action, the most related category would be mistakenly identified. The long-term dependence encoded from actions of this wrong category hinders the prediction of the next action, and even hurts the subsequent recommendation. Consequently, ignoring the uncertainty of the inferred user intent limits the performance of GatedLongRec. On both datasets, when $k=3$, GatedLongRec achieves the best performance. By introducing $k$ types of transition pattern from the inferred user intent enables GatedLongRec to learn more accurate long-term dependence, thus improving the recommendation quality. When $k$ keeps increasing, the transition patterns of more categories are considered as long-term dependence. Consequently the long-term dependence becomes less discriminate even in regard to two actions driven by totally different intents, thus leading to worse recommendations. The above observations demonstrate that the proposed top-$k$ gating network, unlike the hard attention, enhances the performance of GatedLongRec with the appropriate choice of $k$.  

\begin{table}[ht]
\caption{Performance of GatedLongRec with different length of distant actions and recent actions.}
\begin{center}
\begin{tabular}{>{\centering\arraybackslash}p{0.4cm}>{\centering\arraybackslash}p{0.4cm}cc}
    \hline
    \multicolumn{2}{c}{Settings}&\begin{tabular}[c]{@{}cc@{}}\multicolumn{2}{c}{Taobao}\\ \hline
                        Recall@100 & MRR@100
        \end{tabular} & \begin{tabular}[c]{@{}cc@{}}\multicolumn{2}{c}{News}\\ \hline
                        Recall@100 & MRR@100
        \end{tabular} \\
    \hline
    \multirow{4}{*}{M=20} & T=1 & \begin{tabular}[c]{@{}cc@{}} 0.3463 & 0.0944
        \end{tabular} & \begin{tabular}[c]{@{}cc@{}} 0.2167 & 0.0230
        \end{tabular} \\ 
    & T=5 & \begin{tabular}[c]{@{}cc@{}} 0.3608 & 0.0946
        \end{tabular} & \begin{tabular}[c]{@{}cc@{}} 0.2254 & 0.0232
        \end{tabular} \\ 
    & T=20 & \begin{tabular}[c]{@{}cc@{}} 0.3665 & 0.0942
        \end{tabular} & \begin{tabular}[c]{@{}cc@{}} 0.2235 & 0.0234
        \end{tabular} \\ 
    & T=50 & \begin{tabular}[c]{@{}cc@{}} 0.3661 & 0.0948
        \end{tabular} & \begin{tabular}[c]{@{}cc@{}} 0.2351 & 0.0251
        \end{tabular} \\ 
    \hline
     \multirow{4}{*}{T=20} & M=1& \begin{tabular}[c]{@{}cc@{}} 0.3241 & 0.0881
        \end{tabular} & \begin{tabular}[c]{@{}cc@{}} 0.1773 & 0.0185
        \end{tabular} \\ 
     & M=5& \begin{tabular}[c]{@{}cc@{}} 0.3602 & 0.0962
        \end{tabular} & \begin{tabular}[c]{@{}cc@{}} 0.2183 & 0.0221
        \end{tabular} \\ 
     & M=20& \begin{tabular}[c]{@{}cc@{}} 0.3665 & 0.0942
        \end{tabular} & \begin{tabular}[c]{@{}cc@{}} 0.2235 & 0.0234
        \end{tabular} \\ 
     & M=50& \begin{tabular}[c]{@{}cc@{}} 0.3669 & 0.0943
        \end{tabular} & \begin{tabular}[c]{@{}cc@{}} 0.2298 & 0.0241
        \end{tabular} \\ \hline
\end{tabular}
\end{center}
\label{tab:GatedLongRec_length}
\end{table}

% \subsection{Ablation Analysis}

% \textbf{GatedLongRec\_Hard.} As using the category as the supervision introduces uncertainty to the inferred user intent, GatedLongRec changes the hyper-parameter $\lambda$ during training to prevent the user intent from overfitting the category. To analyze how this training schedule affects the performance of GatedLongRec, we develop this variant. During training, this variant sets $\lambda=0.5$. Other parts of this variant is the same as GatedLongRec. 

% \textbf{GatedLongRec\_Hard.} During training, this variant utilizes the ground truth category of the target action to distill the long-term dependence, instead of using the category predicted by the gating network . This variant does not learn to distill the long-term dependence, but distills the long-term dependence in a hard way during training. Other parts of this variant is the same as GatedLongRec. 
% \\
% \textbf{GatedLongRec\_GT.} During training and inference, this variant utilizes the ground truth category of the target action to distill the long-term dependence. Other parts of this variant is the same as GatedLongRec.

\subsection{Long-term Encoder \& Short-term Encoder}
The long-term dependence encoded by the long-term encoder in GatedLongRec is over $T$ distant actions of the predicted categories. To study how $T$ would influence the performance of GatedLongRec, we evaluate our model with different settings of $T$. The result is reported in Table~\ref{tab:GatedLongRec_length}. We can observe that when the number of considered distant actions is small, i.e. $T=1$, the performance of GatedLongRec drops. With too few distant actions, the transition patterns are sparse. Consequently, the model can hardly learn any useful long-term dependence. In addition, when $T$ increases, the introduced gain becomes smaller, especially on Taobao dataset. With the large number of distant actions considered for the long-term dependence in GatedLongRec, the complexity of the structure also grows, which may require more finer-grain or more accurate user intent to contextualize the long-term dependence. 

Likewise, the short-term dependence encoded by the short-term encoder in GatedLongRec is over $M$ recent actions. To understand how the number of recent actions influences the performance of GatedLongRec, we evaluate our model with different values of $M$. From the results in Table~\ref{tab:GatedLongRec_length}, we can observe that when the number of recent actions modeled by the short-term encoder is small, i.e. $M=1$, the performance of GatedLongRec is bad. The transition pattern on very few recent actions is too sparse to be useful, which hurts the modeling of the short-term dependence. In addition, with the small number of recent actions, the user intent representation learnt by the intent encoder would contain little information, thus affecting the modeling of the long-term dependence.

% and also hinders the category prediction by the gating network. GatedLongRec thus can neither accurately capture short-term dependence nor distill the long-term dependence in regard to the target action. 

% In addition, when the number of long-range actions is small, i.e. $T=1$, the performance of GatedLongRec also drops. This is because too few long-range actions limits the modeling of the long-term dependence. These observations show that both the length of short-range actions and the length of long-range actions affect the model's performance; in particular, the length of the short-range actions has higher impact, as it affects both the short-term dependence and the distilled long-term dependence.

To analyze the importance of the long-term dependence and the short-term dependence to the final performance of GatedLongRec, we conduct ablation analysis. We develop two variants of GatedLongRec as follows, \\
\textbf{GatedLongRec\_Short.} This variant only utilizes the short-term encoder to capture the short-term dependence, but excludes other components. The comparison against this variant can show the importance of the long-term dependence.\\
\textbf{GatedLongRec\_Long.} This variant excludes the short-term encoder. Like GatedLongRec, it employs the top-$k$ gating network to structure the long-term dependence in regard to the user intent and utilizes the long-term encoder to capture the long-term dependence. The comparison against this variant can demonstrate the contribution of the short-term encoder. \\

\begin{figure}[t]
\centering
\includegraphics[width=0.45\textwidth]{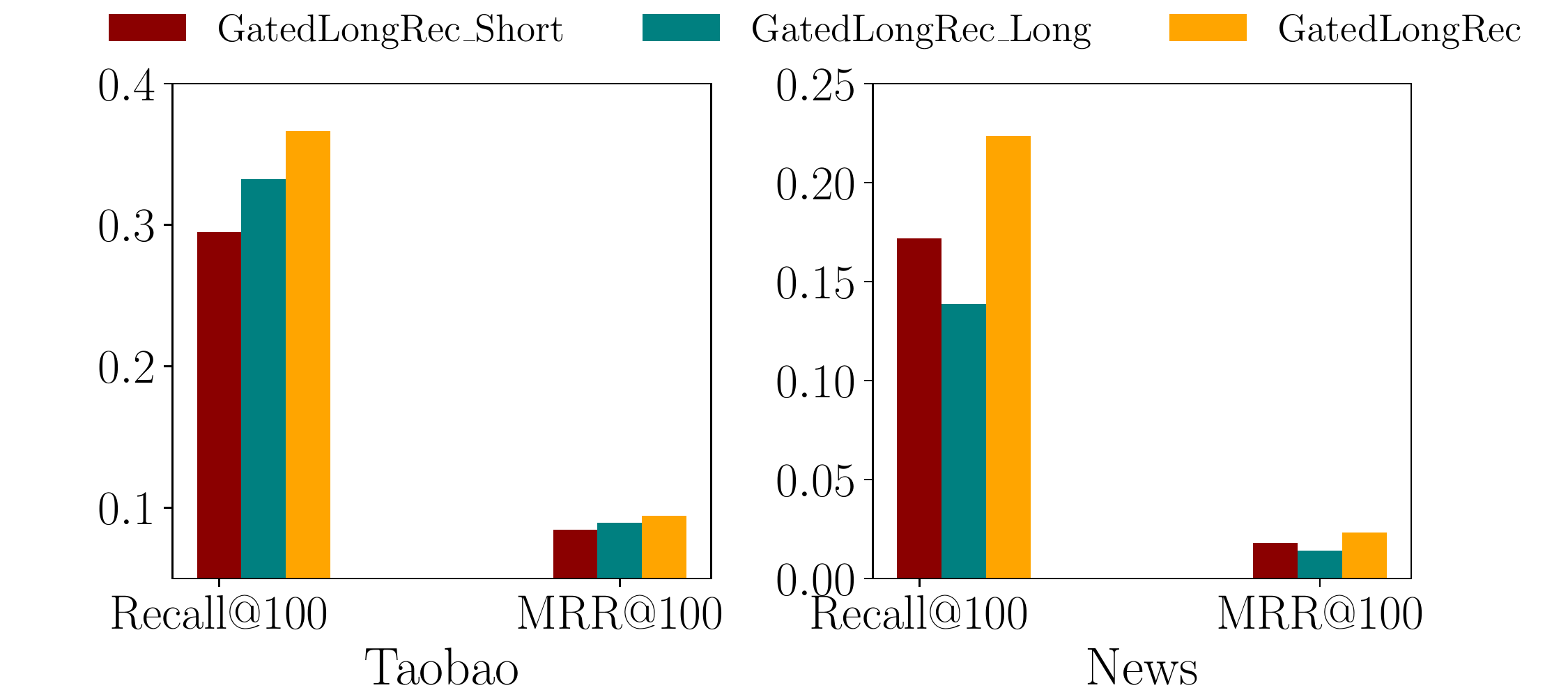} 
\caption{Performance of variants of GatedLongRec for ablation analysis on two datasets.} 
\label{fig:GatedLongRec_ablation}
\end{figure}

The performance of these variants of GatedLongRec is reported in Figure~\ref{fig:GatedLongRec_ablation}. We could observe that GatedLongRec performs much better than GatedLongRec\_Short on both datasets. Structuring the long-term dependence in accordance with the user intent helps GatedLongRec to produce more accurate recommendations, while GatedLongRec\_Short ignores the long-term dependence. Besides, GatedLongRec also outperforms GatedLongRec\_Long, which indicates that the short-term dependence is necessary to make accurate recommendations. 

% Short-term encoder, long-term encoder and gating network are three major components of our model. As we have analyzed how the gating network affects our model in section~\ref{subsub_sec:sec_cate}, we look into how other two components influence the performance via ablation analysis. We develop two variants of our model,  $GatedLongRec\_Long$ which only considers the long-term dependence and $GatedLongRec\_Short$ which only considers the short-term dependence. The comparison between these two variants and our model is as the following table shows. 

\section{Conclusion}
In this paper, we propose a neural sequential recommender, GatedLongRec, which enhances the modeling of the long-term dependence for sequential recommendation. GatedLongRec constructs the structure of the long-term dependence in regard to the proxy user intent of the target action. GatedLongRec learns the latent representation of the proxy user intent by treating the category as the supervision through an intent encoder. Considering the user intent representation is not reliable, GatedLongRec employs a top-$k$ gating network to extract actions of top-$k$ similar categories to the user intent for the long-term dependence modeling. Through a long-term encoder, the transition patterns among actions, i.e., the long-term dependence, is encoded. 

Due to the uncertainty of the proxy user intent representation, GatedLongRec concatenates the short-term dependence learnt by a short-term encoder and the $k$ types of the long-term dependence individually, resulting in $k$ vectors representing the user interest at the target action. GatedLongRec makes predictions based on these $k$ user interest vectors separately and mixes them with the attention weight outputted by the gating network. Extensive experiments on two large datasets demonstrate the necessity of fine-grained modeling of long-term dependence by GatedLongRec for sequential recommendation. 

In this work, the side information, i.e., category is utilized to structure the long-term dependence. It is possible that other kind of information, like text feature, is available and can be used to contextualize the long-term dependence, which we leave as a future work. In addition, we currently only use distance in between the actions to differentiate recent and distant actions; but another important property of users' sequential actions, i.e., action timestamps are not considered yet. It is necessary to incorporate such detailed temporal information to better capture both short- and long-term dependence among actions. As the interactions involve both users and the system, modeling the interactive process through reinforcement learning for the sequential recommendation would be an interesting direction to explore.

% To retrieve the most relevant sequential transition patterns from distant actions, GatedLongRec learns to structure dynamic long-term dependence in regard to the target action. It utilizes a gating network to predict the category of the target action, which serves as a proxy for inferring the ongoing user intent and helps adaptively construct the long-term dependency for the target action. Specifically, the gating network filters the distant actions of the same category to the predicted one of the target action. 
% It then utilizes a long-term encoder to summarize the identified long-term dependency. At the same time, a short-term encoder is used to capture short-term dependency. The latent states of these two types of dependency are used to predict the user's choice of the item at the moment as recommendation. 
% To conquer uncertainty in intent inference, we integrate the top-k most likely category predictions to infer the corresponding long-term dependency structure.
%To mitigate the issue of accumulating errors of the category prediction to the next item prediction, we make item predictions under top-k categories predicted by the gating network and use the mixture of item predictions for recommendations. We jointly optimize tasks of the item and category prediction. 

\bibliographystyle{ACM-Reference-Format}
\bibliography{mybib}

\end{document}